\documentclass[a4paper]{jpconf}
\usepackage{graphicx}
\usepackage[english]{babel}
\begin{document}
\title{Effect of sterile phases on degeneracy resolution capabilities of LBL experiments}

\author{Akshay Chatla and Bindu A. Bambah}

\address{School of Physics, University of Hyderabad, Hyderabad - 500046, India}

\ead{chatlaakshay@gmail.com,bbambah@gmail.com}

\begin{abstract}
In sterile neutrino (3+1) parameterisation, we observe that sterile phases ($\delta_{14},\delta_{24}$) are always together in oscillation probability, even when the MSW effect is considered. We see that the difference between the sterile phases has a more dominating effect over event rates compared to small variations due to changes in individual values. In this work, we show the value of sterile phase difference ($\delta_{14}-\delta_{24}$) that  least affects the parameter degeneracy resolution of $\delta_{13},\theta_{23}$ at long-baseline experiments. 
\end{abstract}

\section{Introduction}
The standard three flavour neutrino mixing is well established. This three-flavor neutrino mixing is parameterized by six parameters, namely three mixing angles - $\theta_{12}$, $\theta_{23}$, $\theta_{13}$, one Dirac CP phase $\delta_{13}$ and two mass squared differences - $\Delta m^2_{31},~\Delta m^2_{21}$ ($\Delta m^2_{ij} = m_i^2 - m_j^2$). The precise measurement of these neutrino oscillation parameters is an important goal for any neutrino oscillations experiments. Sometimes, two different sets of oscillation parameters give rise to same oscillation probability. This causes fake solutions to mask the true solutions which leads to parameter degeneracy. The current unknowns in three flavour model are octant degeneracy of $\theta_{23}$ and mass hierarchy degeneracy (MH) and CP violating phase. Resolving these degeneracies is an important objective for long baseline experiments like NOvA and DUNE.

While, the three flavour model fits with experimental results from most of the experiments well, there are some anomalous results at short baseline (SBL) by LSND \cite{Athanassopoulos:1995iw} and MiniBooNE \cite{AguilarArevalo:2008rc}, which can be interpreted by introducing new  $\mathrm{\Delta}$m$^2 \sim$1 eV$^2$. Since, the number of active neutrinos flavours are limited to three by the LEP \cite{ALEPH:2005ab} experiment, the new neutrino should be a sterile (No weak interaction) neutrino ($\nu_s$). When a one sterile neutrino is added to the three flavour model, it introduces new oscillation parameters (three mixing angles and two CP phases) and is now  called the 3+1 model. These new parameters increase the degrees of freedom and will affect degeneracy resolution capabilities of NOvA and DUNE. In this work, we study the effect of sterile neutrino on degeneracy resolution capabilities of NOvA and DUNE.
\section{Theory}

In this paper, we work on the  3+1 neutrino model i.e., only one light sterile neutrino is added. In 3+1 neutrino model, the neutrino flavour and mass eigenstates are coupled with a 4$\times$4 mixing matrix. A suitable parametrization \cite{Adamson:2017zcg} of the mixing matrix is\begin{equation}\label{eq:3+1}
\mathrm{{U_{3+1}} =  R_{34}(\theta_{34})\tilde{R_{24}}(\theta_{24},\delta_{24})\tilde{R_{14}}(\theta_{14},\delta_{14})\mathbf{R_{23}(\theta_{23})\tilde{R_{13}}(\theta_{13},\delta_{13})R_{12}(\theta_{12})}}.
 \end{equation} 
here $R_{ij}$ and $\tilde{R_{ij}}$ represent real and complex 4$\times$4 rotation in the plane containing the 2$\times$2 sub-block in (i,j) sub-block\\\\
\begin{equation}
R_{ij}^{2\times 2}= \left( \begin{array}{cc}
c_{ij} & s_{ij} \\
-s_{ij} & c_{ij} \end{array} \right) \quad\quad 
\tilde{R_{ij}}^{2\times2}= \left( \begin{array}{cc}
c_{ij} & \tilde{s_{ij}} \\
-\tilde{s_{ij}}^* & c_{ij} \end{array} \right) \\\\ \end{equation}
Where, $ c_{ij} = \cos\theta_{ij}$, $ s_{ij} = \sin\theta_{ij}, \tilde{s_{ij}}=s_{ij}e^{-i\delta_{ij}}$ and $\delta_{ij} $ are the CP phases.\\

The bold matrices in eqn.\ref{eq:3+1} represent standard 3 flavour model. We see that addition of one sterile neutrino introduces 3 new mixing angles ($\theta_{14},\theta_{24},\theta_{34}$) and 2 new CP-phases ($\delta_{14},\delta_{24}$). The measurement of these new parameters is important in the study of sterile neutrinos. Short-baseline(SBL) experiments which are sensitive to oscillations lead by new $\Delta \rm{m}^{2} \simeq 1$eV$^2$ are good places to measure sterile-mixing angles. Although SBL experiments can give good bounds on sterile mixing angles, they are not sensitive to new CP-phases introduced by $\nu_s$ as they need longer distances to become measurable. So, experiments like NOvA and DUNE will be useful to measure these parameters. The electron neutrino appearance probability, P$_{\mu e}$ for LBL experiments in the 3+1 model for vacuum, after averaging $\Delta$m$^2_{41}$ oscillations, can be expressed as a sum of the four terms based on the CP terms they contain \cite{Chatla:2018sos}

\begin{equation} \label{eq.4}
P_{\mu e} ^{4\nu} \simeq P_1 + P_2(\delta_{13}) + P_3(\delta_{14}-\delta_{24})+P_4(\delta_{13}-(\delta_{14}-\delta_{24})).
\end{equation}

The last two terms of eq.\ref{eq.4}, give the sterile CP phase dependence terms. The term  P$_{3}(\delta_{14}-\delta_{24})$ depends on the sterile CP phases $\delta_{14}$ and $\delta_{24}$, while P$_{4}$ depends on a combination of $\delta_{13}$ and $\delta_{14}-\delta_{24}$. We see that the sterile CP phases persist in the P$_{\mu e}$ even after averaging out $\Delta$m$^2_{41}$ lead oscillations. So, experiments like NOvA and DUNE will be effected by sterile phases, thus effecting their parameter degeneracy resolution capacity. We also note that the sterile phases $\delta_{14}$ and $\delta_{24}$ are always together as $\delta_{14}-\delta_{24}$ in a linear dependent form. From this linear dependency, we can deduce that only the difference between the sterile phases ($\delta_{14}-\delta_{24}$) effect the oscillation probability in vacuum case.

Now, we check whether individual values of  $\delta_{14}$ and $\delta_{24}$ cause any impact on oscillation probabilities after matter effects \cite{Smirnov:2004zv} are introduced. GLoBES (General Long Baseline Experiment Simulator)\cite{Huber:2004ka,Huber:2007ji} which  is a software package for the simulation of long baseline neutrino oscillation experiments is used for simulating NOvA and DUNE while including matter effects. The experimental details we used for the simulation are in Ref.\cite{Adamson:2016tbq,Alion:2016uaj}. The 2019 NOvA analysis\cite{Acero:2019ksn}, gave two best fit points for normal hierarchy-higher octant(NH-HO) and normal hierarchy-lower octant(NH-LO). We take these best fit points as the true values for our analysis. We take upper bounds given in NOvA\cite{Adamson:2017zcg} neutral current analysis paper for sterile mixing angles. The
mass squared difference was taken fixed at $\Delta m^2_{41} =1$eV$^2$.

\begin{figure}[ht]
\begin{center}
    \includegraphics[width=1.0\columnwidth,height=18pc]{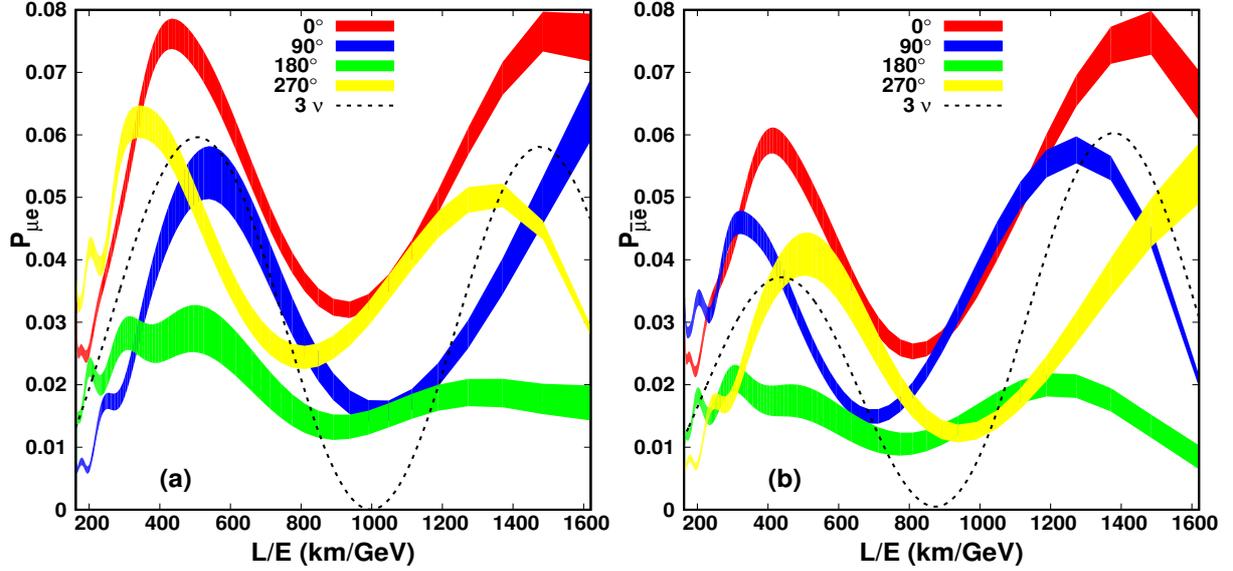}
\caption{\label{fig1} (a)The electron neutrino((b)The electron anti-neutrino) appearance probability P$_{\mu e}$(P$_{\overline{\mu e}}$) for NOvA as function of L/E(km/GEV) for various bands of ($\delta_{14}-\delta_{24}$) while varying individual values of $\delta_{14},\delta_{24}$ inside the band. }
\end{center}
\end{figure}

\begin{figure}[ht]
\begin{center}
  \includegraphics[width=1.0\columnwidth,height=18pc]{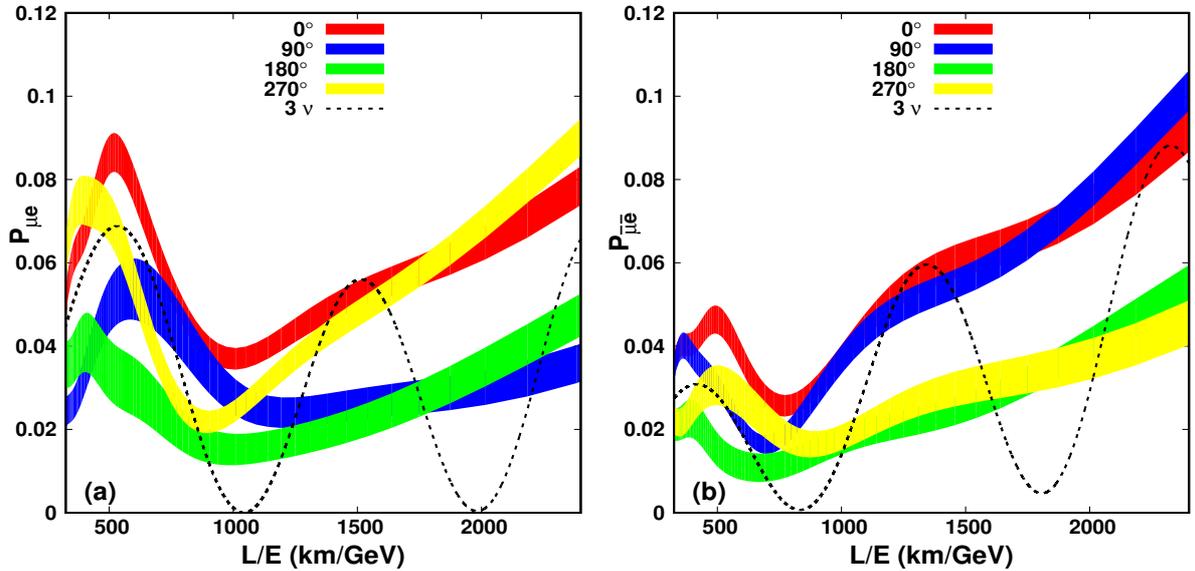}
\caption{\label{fig2} (a)The electron neutrino((b)The electron anti-neutrino) appearance probability P$_{\mu e}$(P$_{\overline{\mu e}}$) for DUNE as function of L/E(km/GEV) for various bands of ($\delta_{14}-\delta_{24}$) while varying individual values of $\delta_{14},\delta_{24}$ inside the band. }
\end{center}
\end{figure}

In Fig:\ref{fig1}(a), we plot P$_{\mu e}$ for  the NOvA experiment with the inclusion of matter effects using GLoBES. Various bands of ($\delta_{14}-\delta_{24}$) are plotted while varying individual values of $\delta_{14},\delta_{24}$ inside the band. In Fig:\ref{fig1}(b), we plot the  same type of plot for electron anti-neutrino appearance.  We note that  even after considering matter effects, individual values of $\delta_{14},\delta_{24}$ cause only small variations in P$_{\mu e}$(P$_{\overline{\mu e}}$), while the difference $\delta_{14}-\delta_{24}$ still plays a major role. In Fig:\ref{fig2}(a,b), we plot similar plots for DUNE. We see that for the DUNE case also, only the difference $\delta_{14}-\delta_{24}$ is important. Thus, we can say that the difference between the sterile phases has a more dominating effect on oscillation probability compared to small variations due to changes in individual values. So, in this work, we find which value of  $(\delta_{14}-\delta_{24})$, least affects on the parameter degeneracy of resolution of $\delta_{13},\theta_{23}$ for NOvA and DUNE.

\section{Results}
We show allowed regions in sin$^2 \theta_{23}$-$ \delta_{cp}$ plane from NOvA and DUNE simulation data for different ($\delta_{14}-\delta_{24}$) values taking 2019 NOvA results\cite{Acero:2019ksn} as true values. We marginalise over all the sterile mixing angles and $\Delta m^2_{31}$ in our analysis. 

\begin{figure}[ht]
\begin{center}
  \includegraphics[width=1.0\columnwidth,height=16pc]{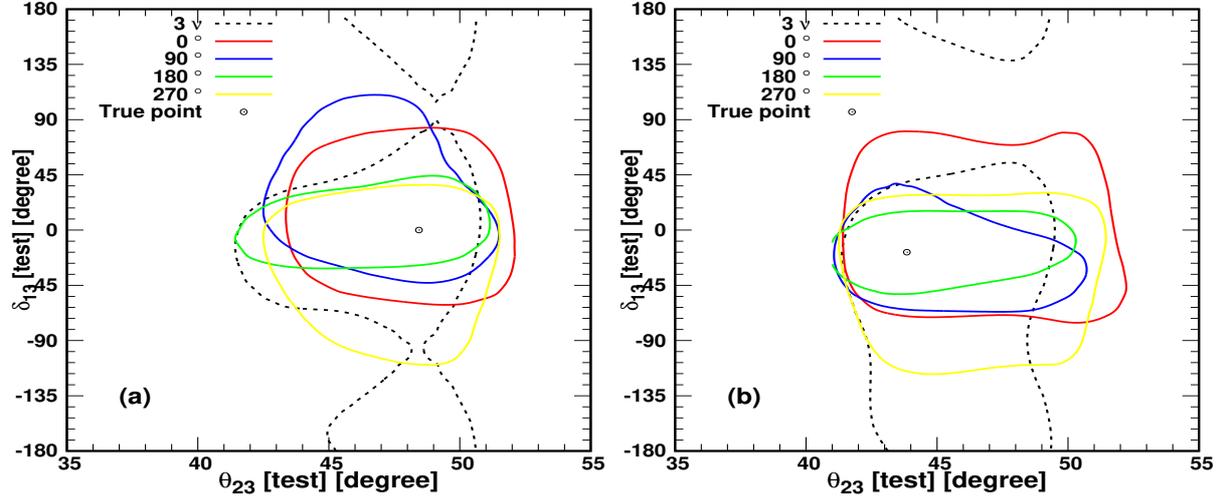}
\caption{\label{fig3} Contour plots of allowed regions in the test plane, $\theta_{23}$ vs $\delta_{13}$ for different $(\delta_{14}-\delta_{24})$ values at 90$\%$ C.I regions for NOvA[3+$\overline{3}$]. We take NH-HO(NH-LO) as true values for a(b) case. }
\end{center}
\end{figure}

In figure \ref{fig3}, we show allowed areas at 90$\%$ C.I for NOvA[3+$\bar{3}$] i.e., 3 years of neutrino and 3 years of anti-neutrino run time for NH-HO and NH-LO true values. The figure \ref{fig3}a, in which NH-HO is taken as true values, we observe that the parameter resolution sensitivity is highest when value of $\delta_{14}-\delta_{24}=180^\circ$. Same case follows for figure \ref{fig3}b, in which NH-LO is taken as true values. 

\vspace{-2em}

\begin{figure}[ht]
\begin{center}
  \includegraphics[width=1.0\columnwidth,height=16pc]{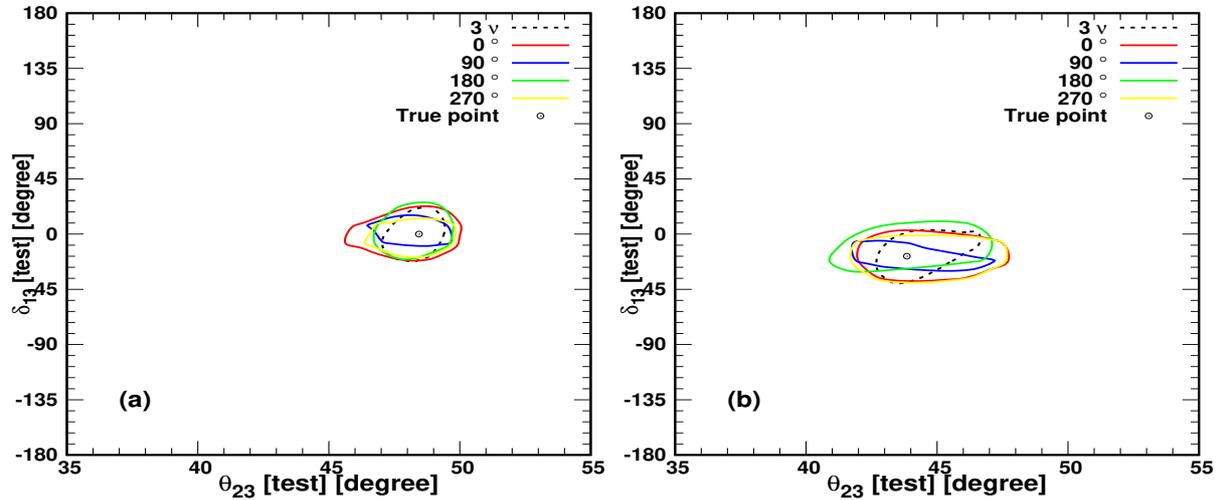}
\caption{\label{fig4} Contour plots of allowed regions in the test plane, $\theta_{23}$ vs $\delta_{13}$ for different $(\delta_{14}-\delta_{24})$ values at 90$\%$ C.I regions for DUNE[3+$\overline{3}$]. We take NH-HO(NH-LO) as true values for a(b) case. }
\end{center}
\end{figure}

In figure \ref{fig4}, we show allowed areas at 90$\%$ C.I for DUNE[3+$\bar{3}$] i.e., 3 years of neutrino and 3 years of anti-neutrino run time for NH-HO and NH-LO true values. The figure \ref{fig4}a, in which NH-HO is taken as true values, we observe that the parameter resolution sensitivity is highest when value of $\delta_{14}-\delta_{24}=90^\circ$. Same case follows for figure \ref{fig4}b, in which NH-LO is taken as true values. 

In summary, we find that only the difference between sterile phases $(\delta_{14}-\delta_{24})$ is important, where as the individual values of $\delta_{14},\delta_{24}$ play a negligible role in case of neutrino oscillation at LBL experiments. We note the NOvA and DUNE have different value of $(\delta_{14}-\delta_{24})$, at $180^\circ$ for NOVA and $90^\circ$ for DUNE at which they get good parameter resolution. This difference in the result maybe due to difference in matter effects and statistics between NOvA and DUNE. This is our subject for future investigation. More work is needed to be done to include different run-times of NOvA and DUNE.

\end{document}